\documentclass[psfig,useAMS,usenatbib,letterpaper]{mn2e}
\usepackage{aas_macros}
\usepackage{psfig}
\usepackage{graphicx}
\usepackage{multirow}
\title[Noise bias in lensing shape measurement]{Noise bias in weak lensing shape measurements} 
\author[Refregier et al.]{Alexandre Refregier,$^{1}$
  Tomasz Kacprzak,$^2$ Adam Amara,$^{1}$ 
Sarah Bridle,$^{2}$  \newauthor Barnaby Rowe$^{2,3,4}$ \\
$^{1}$Institute for Astronomy, ETH Zurich, Wolfgang-Pauli-Strasse 27, CH-8093 Zurich, Switzerland\\
$^{2}$Department of Physics \& Astronomy, University College London,
Gower Street, London, WC1E 6BT\\
$^3$Jet Propulsion Laboratory, California Institute of Techonology,
4800 Oak Grove Drive, Pasadena, CA 91109, USA\\
$^4$California Institute of Technology, 1200 East California Boulevard,
Pasadena, CA 91106, USA}
\date{\today}

\begin{document}

\maketitle

\begin{abstract}
  Weak lensing experiments are a powerful probe of cosmology through
  their measurement of the mass distribution of the universe. A
  challenge for this technique is to control systematic errors that
  occur when measuring the shapes of distant galaxies. In this paper
  we investigate noise bias, a systematic error that {arises} from
  second order noise terms in the shape measurement process.  We first
  derive analytical expressions for the bias of general Maximum
  Likelihood estimators {(MLEs)} in the presence of additive noise.
  We then find analytical expressions for a simplified toy model in
  which galaxies are modeled and fitted with a Gaussian with its size
  as a single free parameter. Even for this very simple case we find a
  significant effect.  We also extend our analysis to a more realistic
  6-parameter elliptical Gaussian model.  We find that the noise bias
  is generically of the order of the inverse-squared signal-to-noise
  ratio (SNR) of the galaxies and is thus of the order of a percent
  for galaxies of SNR of 10, i.e.\ comparable to the weak lensing
  shear signal. This is nearly two {orders} of magnitude {greater}
  than the systematics requirements for future all-sky weak lensing
  surveys. We discuss possible ways to circumvent this effect,
  including a calibration method using simulations discussed in an
  associated paper.
\end{abstract}


\section{Introduction}
Weak gravitational lensing is a technique to map the distribution of
dark matter in the universe \citep[see e.g.][for
reviews]{refregier03,hoekstraj08}.  It relies on measurement of the
apparent shapes of distant galaxies that are distorted due to matter
inhomogeneities along the line of sight. Weak lensing offers great
prospects for the measurement of cosmological parameters
\citep{esoesa,detf}.  In particular, the measurements of dark energy
parameters with future wide field surveys is very promising but places
strong requirements on weak lensing measurements and in particular in
the control of systematics.

The main potential systematic effects are generally considered to be:
(i) galaxy shape measurement from galaxy images; (ii) galaxy distance
measurement using photometric redshifts; (iii) galaxy intrinsic
alignments arising from the galaxy formation process; (iv) accuracy of
theoretical predictions of dark matter clustering.  We focus on the
first of these in this paper.

In most cases the gravitational lensing effect produces a matrix
distortion stretching of the galaxy image. This image shear must be
uncovered in the presence of nuisance observational effects,
including: image blurring due to the atmosphere and telescope optics;
image pixelisation due to the nature of photon detectors; and noise
due to the finite number of photons from the galaxy and other
backgrounds. Furthermore, the intrinsic properties of the galaxy prior to
lensing distortion are unknown. 

The first detection of this shearing effect was made
by~\citet{tysonwv90} and repeated by~\citet{bonnetmf94} who also
developed methods for removing the image convolution effects. This was
taken to a new level by~\citet{kaisersb95} in a method that is widely
referred to as KSB and that has remained the most widely used shear
measurement method to this day. Essentially the KSB method uses
weighted quadrupole moments of images to calculate shears, and
corrects the shears for the weighting function. This was further
improved in ~\citep{kaiser00}. An alternative approach using a simple
galaxy model to forward fit the data was proposed in~\citet{kuijken99}
and implemented in \citet{bridlekbg02} and \citet{millerkhhv07}. More
flexible models using Gauss-Laguerre polynomials, or shapelets, have
also been proposed \citep{bernsteinj02,refregierb03}. Each of these
approaches has potential strengths and drawbacks. For instance the
limitations of model-fitting methods {were} explored in
\citet{melchioretal10} and \citet{voigtb10}, and potentially mitigated
by \citet{bernstein10}.

There have been several {simulation} challenges to assess how well
current methods can measure gravitational shear and to encourage
development of new methods.  The Shear TEsting Program (STEP) 1
Challenge provided a suite of simulated images using relatively simple
galaxy models but a {realistic} image blurring model. The galaxies
were distributed with random positions across the image and the same
shear was used to distort every galaxy in a given large image. It was
found that the existing methods that had already been applied to
observational data were sufficiently good to merit the science results
on that data~\citep{STEP1mnras}.

The STEP2 Challenge used more realistic galaxy models and a wider
range of blurring models~\citep{STEP2mnras} and reached similar
conclusions despite this additional complexity. However, neither
challenge was sufficiently large to forecast the efficacy of existing
methods for use on future surveys, and neither challenge was able to
address \emph{all} potential sources of measurement bias in real data
(such as uncertainty about the image Point Spread Function or PSF;
see, e.g., \citealp{paulinetal08,paulinetal09,rowe10}).  In addition,
while it was possible in many cases to positively detect biases in
weak lensing shape measurement methods, due to the complexity and
realism of the STEP challenges it was not always possible to attribute
definite causes for these effects.  In the complex, multi-stage
analysis required for weak lensing measurement it can be very
difficult to isolate individual causes of systematic bias, and yet
diagnosing these individual contributions is an important ongoing
process in the development of accurate measurement methodology.

The GRavitational lEnsing Accuracy Testing 2008 (GREAT08) Challenge
was much simpler: it reverted to simpler galaxy models; avoided
overlapping galaxies; and used similar properties for all galaxies in
a large image~\citep{GREAT08mnras}. It was designed to attract new
methods from outside the weak lensing community, in particular from
computer scientists. Most significantly, the number of galaxies was
chosen to test methods at the level required for surveys in the
foreseeable future, as calculated in~\citet{Amara:2007as}.  A new
approach won the competition, inspired by~\citet{kuijken99}, which
took advantage of the fact that the same shear was used for many
galaxies at a time, by `stacking' the galaxies
\citep{lewis09,hosseini2009}. Although progress has been
  substantial, questions still remain about the likely issues that
need to be overcome to reach the precision needed for future all-sky
surveys.

In this paper we study noise bias, one of the systematic effects that
can affect weak lensing measurements.  It arises from high-order noise
terms in the measurement of the shape parameters of galaxies,
increasing in magnitude at low galaxy signal-to-noise ratio (SNR).
Its effects on second order moment measurements from convolved
Gaussian galaxy images has been described by \citet{hirataetal04}.

To study this effect in a forward-fitting weak lensing measurement
context we first derive general expressions for the variance and bias
of Maximum-Likelihood Estimators (MLE) of model parameters in the
presence of additive Gaussian noise (\S\ref{general}). We then apply
it to a one-parameter toy model consisting of the Maximum-Likelihood
  (ML) fitting of the size of a Gaussian galaxy model to a Gaussian
galaxy convolved with a known Gaussian PSF (\S\ref{gaussian}). While
this model is clearly oversimplified it illustrates the principle of
noise bias and its amplitude.  We then extend this result by
considering a more realistic model consisting of an elliptical
Gaussian galaxy with 6 free parameters (\S\ref{elliptical}). In
\S\ref{conclusions}, we discuss the consequences of our findings and
possible remedies, and summarise our conclusions, including a
calibration method using simulations discussed in an associated paper
(Kacprzak et al. 2012, hereafter K12).

\section{General 2D shape estimation}
\label{general}
In this section, we study the general problem of the estimation of the
shape parameters of a 2D object in the presence of additive,
uncorrelated Gaussian noise. For weak lensing these results are
applicable to the {measurement of galaxy shapes, and to the estimation
  of the instrument PSF using stars in the image}.  The general
analytical results that we derive {will serve as a useful base for
  comparison with} the more realistic conditions studied by K12 {using
  numerical simulations}.

\subsection{General results}
Let us thus consider the observed 2D surface brightness $f_{\rm
  obs}({\mathbf x})$ of an object that is described by a model
$f({\mathbf x};{\mathbf a})$, where ${\mathbf x}$ is the position on
the image and ${\mathbf a}$ is the vector of parameters describing the
shape of the object. We can write the observed surface brightness as
\begin{equation}
f_{\rm obs}({\mathbf x})=f({\mathbf x};{\mathbf a}^{\rm t})+n({\mathbf x})
\label{eq:f_obs}
\end{equation}
where ${\mathbf a}^{t}$ are the true shape parameters of the object
and $n({\mathbf x})$ is the noise which is assumed to be uncorrelated
and Gaussian with $\langle n({\mathbf x}) \rangle=0$ and $\langle
n({\mathbf x})^2 \rangle= \sigma_n^2$.

With these assumptions the log likelihood of the data given the model
is $\ln L = - \chi^2/2$, where the usual $\chi^2$-functional is given
by
\begin{equation}
  \chi^2({\mathbf a})= \sum_{p} \sigma_{n}^{-2} \left[ f_{\rm obs}({\mathbf x}_p) - f({\mathbf x}_p;{\mathbf a}) \right]^{2},
\label{eq:chi2}
\end{equation}
where the sum is over all pixels $p$ in the image. 

The {MLE} $\hat{\mathbf a}$ for the shape parameters of the object can
then be constructed by requiring that $\chi^2$ is minimised at
${\mathbf a}=\hat{\mathbf a}$.  MLEs were first studied by
\citet{fisher1922}, then later by \citet{rao1973} and
\citet{cramer1999}. They are commonly used estimators in statistics
and have several desirable properties, including consistency, which
{requires} that in the limit of high SNR the MLE recovers the true
values {$\textbf{a}^{\rm t}$} of the estimated parameters.

In Appendix~\ref{app:general} we derive general properties for the
{MLE} $\hat{\mathbf a}$ using an expansion in the inverse SNR {of the
  object. There and in what follows we label SNR using the parameter
  $\rho $}. We first show that the covariance of the estimated
parameters is, to leading order, given by
\begin{equation}
  \label{eq:cov}
  {\rm cov}[\hat{a}_i,\hat{a}_j]=(F^{-1})_{ij} + { O(\rho^{-4}),}
\label{eq:covariance}
\end{equation}
where the Fisher matrix is given by
\begin{equation}
\label{eq:fisher}
F_{ij}=  \sum_p \sigma_n^{-2} \frac{\partial f({\mathbf x}_p;{\mathbf a}^{t})}{\partial a_i}\frac{\partial f({\mathbf x}_p;{\mathbf a}^{t})}{\partial a_j}{.}
\end{equation}

We also find that the bias in the parameters $b[\hat{a}_i]=\langle
\hat{a}_i \rangle - a_i^{\rm true}$ is given by
\begin{equation}
\label{eq:bias}
\label{bias}
b[\hat{a}_i]=-\frac{1}{2}(F^{-1})_{ij} (F^{-1})_{kl} B_{jkl} + { O(\rho^{-4})}  ,
\end{equation}
where the summation convention over repeated indices is assumed and
the bias tensor is given by
\begin{equation}
B_{ijk}= \sum_p \sigma_n^{-2} \frac{\partial f({\mathbf x}_p;{\mathbf a}^{\rm t})}{\partial a_i}
\frac{\partial^2 f({\mathbf x}_p;{\mathbf a}^{\rm t})}{\partial a_j \partial a_k}.
\label{eq:bias_tensor}
\end{equation}

It is often useful to consider functions $g_i({\mathbf a})$ of the
parameters. The {covariances} of these functions are given by
\begin{equation}
\label{eq:g_cov}
{\rm cov}[g_i,g_j]= \frac{\partial g_i}{\partial a_k} \frac{\partial g_j}{\partial a_l} {\rm cov}[a_k,a_l],  
\end{equation}
while their bias is given by
\begin{equation}
\label{eq:g_bias}
b[g_i]=\frac{\partial g_i}{\partial a_k} b[a_k]{.}
\end{equation}

\subsection{Properties}
\label{properties}
As can be seen from Equation~\ref{eq:bias}, the bias tensor depends on
second order derivatives of the model $f({\mathbf x},{\mathbf a})$ in
the parameters ${\mathbf a}$ and therefore vanishes for linear models.
This restates the known fact that MLEs may be biased in general,
except in the case of linear models.  As noted in
Appendix~\ref{app:general}, the present bias arises from second order
noise terms and is therefore {referred} to as `noise bias'.

We then note that the squared error in the parameters
$\sigma^2[a_i]={\rm cov}[a_i,a_i]$ and the bias $b[a_i]$ are of order
\begin{equation}
b[\hat{a}_i]/a_i^t \sim [\sigma[\hat{a}_i]/a_i^t]^2  \sim \rho^{-2},
\label{eq:scalings}
\end{equation}
in dimensionless units. In the limit of high {SNR}, $\rho \rightarrow
\infty$, both tend to zero so that the estimator tends to the true
value $\hat{\mathbf a} \rightarrow {\mathbf a}^t$, thus recovering the
consistency property of MLEs (e.g.\ \citealt{cramer1999}).

For finite values of $\rho$ the statistical error and bias of the
parameters can be non negligible. Weak lensing shape measurements are
typically performed down to $\rho \sim 10$ to maximise the surface
density of galaxies. In this case, the statistical RMS error will be
of order $\rho^{-1} \sim 0.1$ which is consistent with the typical
observed shape noise per galaxy of about $\delta \gamma \sim 0.3$, and
which {includes} this statistical measurement error and the
distribution of the intrinsic shape of galaxies. The bias in the
parameters in this regime will be of order $\rho^{-2} \sim 0.01$ which
is comparable to the weak lensing shear signal $\gamma \sim {0.02}$
and may contribute to explain why some methods do not perform better
\citep[e.g.][]{GREAT08mnras}.  As shown by \citet{Amara:2007as}, the
requirement for the variance of the shear systematics is
$\sigma^2_{\mathrm sys} \sim 10^{-7}$ which corresponds to the
systematic shear error of $\delta \gamma \sim 3\times 10^{-4}$.  This
is almost two orders of magnitude smaller than predicted by the
current analysis {of noise bias}.

We also note that the expressions for the variance and bias of the ML
estimators are expressed in Equations~\ref{eq:covariance} and
\ref{eq:bias} in terms of a sum over pixel positions, but can {often
  be} more conveniently evaluated in the continuum limit where the
pixel size is small compared to the object size. This approximation is
given by Equation~\ref{eq:continuous}, {re-expressing the sum} as an
integral {over} the 2D image.

Seemingly counter-intuitively, the bias of the derived parameters
$g_i(\hat{\mathbf a})$ is not equal in general to the bias that would
be derived had it instead been chosen to find the MLE of the
parameters $\hat{g}_i$ directly. This can be understood from examining
the covariance transformation rule described in Equation
\ref{eq:g_cov}.  Thus, the exact value of the bias {for any parameter
  of interest may} depend on the parametrisation of the model
{itself}.  We will show an example of this property in the following
simplified example.

\section{Circular Gaussian model}
\label{gaussian}
To illustrate the above results, we first consider the case of the
measurement of the size of a 2D, circular, Gaussian galaxy without any
other free parameters. This is a highly simplified illustration of the
shape measurement problem and should therefore be considered as a toy
model that captures the main features of the effect of noise bias.

\subsection{Case without PSF convolution}
\label{1dgaussian_nopsf}
First, let us consider the case where the galaxy is not convolved with
the PSF of the instrument. In this case the galaxy surface brightness
is given by
\begin{equation}
f({\mathbf x};a)= f_{0}  \exp \left[ -\frac{r^2}{2 a^2} \right],
\end{equation}
where $r^2=x_1^2+x_2^2$, $f_0$ is a a normalization controlling the
flux of the galaxy, and $a$ is the rms size.

To characterise the {SNR} $\rho$ of the galaxy, we temporarily
consider $f_0$ as the free parameter while keeping $a$ fixed. Using
the continuum limit (Eq.~\ref{eq:continuous}), we can analytically
integrate Equation~\ref{eq:covariance} for the variance of the
estimator for $f_0$ and obtain
\begin{equation}
\rho[f_0] = \frac{f_0}{\sigma[f_0]} = \frac{\sqrt{\pi} f_0 a}{\sigma_n h},
\end{equation}
where $h$ is the pixel scale and $\sigma_n$ is the noise rms as in
\S\ref{general}. Here and in the following, we drop the $~\hat{}~$ and
$^{t}$ symbols to simplify the notation when it does not lead to
ambiguities.  This definition of the SNR can be considered as the
ideal detection signal-to-noise of the galaxy, corresponding to a
perfect knowledge of the galaxy shape and position, or equivalently to
an ideal matched filter.

Now, considering $a$ as the only free parameter (and thus leaving
$f_0$ fixed) we again integrate Equation~\ref{eq:covariance}
analytically in the continuum limit and obtain
\begin{equation}
\frac{\sigma[a]}{a} = \frac{1}{\sqrt{2}} \rho[f_0]^{-1} + O(\rho^{-2}),
\end{equation}
which scales as $\rho^{-1}$ as noted in
\S\ref{properties}. Integrating Equation~\ref{eq:bias}, we find that
the bias in this case vanishes at second order, i.e.\ $b[a] = 0 +
{O(\rho^{-4})}$. This is due to a cancelation which we find occurs for
any 2D circular galaxy model which can be written as
\begin{equation}
f({\mathbf x};a) = f_0 \phi(r/a),
\label{eq:symmetry}
\end{equation}
where $\phi$ is any function describing the galaxy
profile. Interestingly, this cancelation only occurs in two
dimensions, and the second order bias term does not vanish in $1$ or
$>2$ dimensions even if the above scaling symmetry holds.

\subsection{Case with PSF convolution}
Let us now consider the case of interest in practice where the
circular Gaussian galaxy is convolved with a PSF due to the instrument
and the atmosphere. In the spirit of the toy model, we make the
simplifying assumption that the PSF is itself circular and Gaussian
with an rms size $p$. Since the convolution of two Gaussians is
{another} Gaussian with standard deviations adding in quadrature, the
model in this case is
\begin{equation}
f({\mathbf x};a)= f_{0}  \exp \left[ -\frac{r^2}{2 (a^2+p^2)} \right],
\end{equation}
where $f_0$ is a normalization controlling the flux of the galaxy.
The PSF size $p$ is assumed known and the noise {assumed to be}
Gaussian with an rms of $\sigma_n$.

In this case, the ideal SNR defined as in \S\ref{1dgaussian_nopsf}
becomes
\begin{equation}
\rho = \frac{f_0}{\sigma[f_0]} = \frac{\sqrt{\pi} f_0 \sqrt{a^2+p^2}}{\sigma_n h},
\end{equation}
and the {uncertainty} in the MLE for $a$ is given by
\begin{equation}
\frac{\sigma[a]}{a} = \frac{1}{\sqrt{2}} \left[ 1 + \left( \frac{p}{a} \right)^2 \right] \rho[f_0]^{-1}
+ O(\rho^{-2}).
\end{equation}
Because the presence of the PSF breaks the scaling symmetry of
Equation~\ref{eq:symmetry}, the bias does not vanish to second order
and we find
\begin{equation}
\frac{b[a]}{a}  = - \frac{1}{4}  \left[ 1 + \left( \frac{p}{a} \right)^2 \right] \left( \frac{p}{a} \right)^2
\rho[f_0]^{-2} +  {O(\rho^{-4})}.
\end{equation}
We note that we recover the results of \S\ref{1dgaussian_nopsf}
without a PSF when we set $p=0$ in the expressions above.  We also see
that the scalings of Equation~\ref{eq:scalings} hold for the rms error
and bias with pre-factors that depend on the ratio of the galaxy to
PSF size.

We also notice that if we had instead estimated the convolved galaxy
parameters directly, and obtained the deconvolved size as a derived
parameter (using Equation~\ref{eq:g_bias}), the bias would then have
vanished to this order.  But this would also allow unphysical values
of the parameters, with $a^2<0$.  This is an illustration of the fact
that the bias of physical parameters depends in general on the
specific parametrisation of the estimated model, as discussed in
\S\ref{properties}.

\subsection{Simulations}
In order to check the validity of the expansion described in
\S\ref{general} and Appendix \ref{app:general}, and gain insights in
the origin of the bias, we performed numerical simulations of this toy
model. We considered a range of SNR for a circular Gaussian galaxy of
true size $a^t=4$ convolved with a circular Gaussian PSF of size
$p=5.33$ pixels.  This corresponds to a ratio of the convolved galaxy
size to the PSF size of $\sqrt{a^{t2}+p^2}/p \simeq 1.25$ which is
typically used as the limit for weak lensing surveys.

Even for this very simple {one}-parameter toy model, we find that
great care must be taken for the implementation of the minimization of
the $\chi^2$ function. Readily available minimiser algorithms produced
{artifacts} that appear to depend on the choice of algorithm at the
high level of precision required for weak lensing. For the present
simulations, we computed $\chi^2(a)$ on a grid in the interval
$0.1<a<10$ pixels with a grid size of $\Delta a =0.0033$ pixels and
found the minimum by direct search.

\begin{figure}
\begin{center}
\includegraphics[width=80mm]{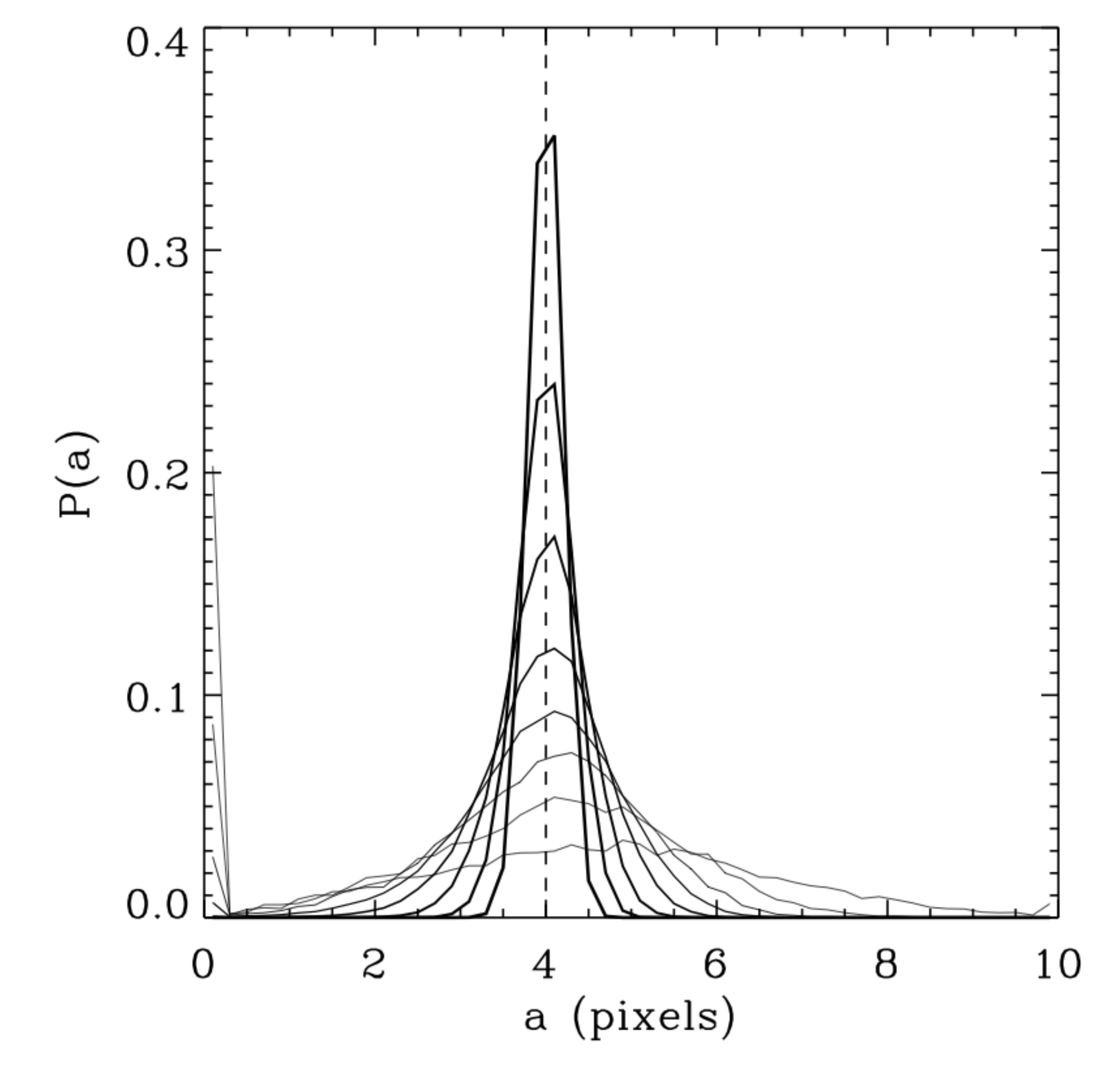}
\caption{Distribution $P(a)$ of the MLE estimator for the size $a$ of
  the 2D Gaussian from repeated realisations. The curved from broad
  (thin lines) to sharp (thick lines) correspond to SNR $\rho[f_0]$ of
  3, 5, 7, 9, 12, 17, 25, 40, respectively. The true value of the
  parameter $a^t=4$ pixels is shown as the vertical dashed line. The
  PSF size is $p=5.33$ pixels corresponding to a convolved galaxy size
  to the PSF size of $\sqrt{a^{t2}+p^2}/p \simeq 1.25$.}
\label{fig:pdf_a}
\end{center}
\end{figure}

Figure~\ref{fig:pdf_a} shows the resulting Probability Distribution
Function (PDF) $P(a)$ of the estimator for $a$ for a range of SNR
$\rho[f_0]$. At high SNR, the PDF is nearly Gaussian and peaks close
to the true value $a^{t}=4$ pixels. As the SNR decreases, the PDF main
peak shifts towards the right, while a secondary peak at $a=0$ starts
developing. The complicated combination of these two effects
contribute to the dependence of the bias and rms error on SNR.

\begin{figure}
\begin{center}
\includegraphics[width=80mm]{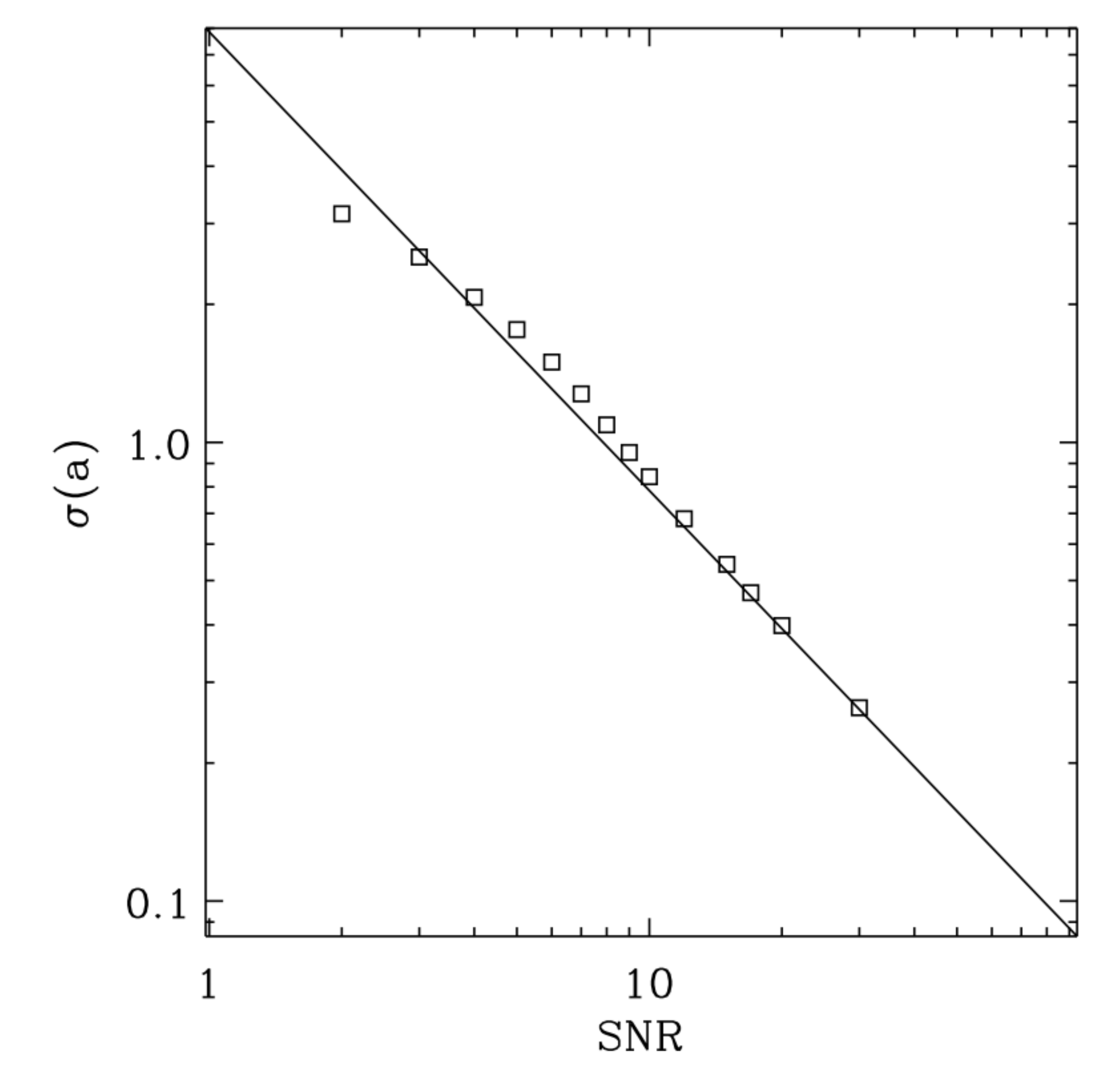}
\caption{Standard deviation $\sigma[a]$ of the size estimator $a$ as a
  function of signal-to-noise ratio ${\rm SNR}=\rho[f_0]$. The
  expectation from the analytical prediction (solid line) is compared
  to the measurements from repeated experiments.}
\label{fig:sigma_a}
\end{center}
\end{figure}

Figure~\ref{fig:sigma_a} shows the dependence of the rms variance
$\sigma[a]$ on the SNR $\rho[f_0]$ . We see that our expression in
Equation~\ref{eq:covariance} is a good approximation for $\rho[f_0]
\ga 10$. The deviations below this value are not surprising since
higher order terms are expected to become important in the low SNR
limit.

\begin{figure}
\begin{center}
\includegraphics[width=70mm]{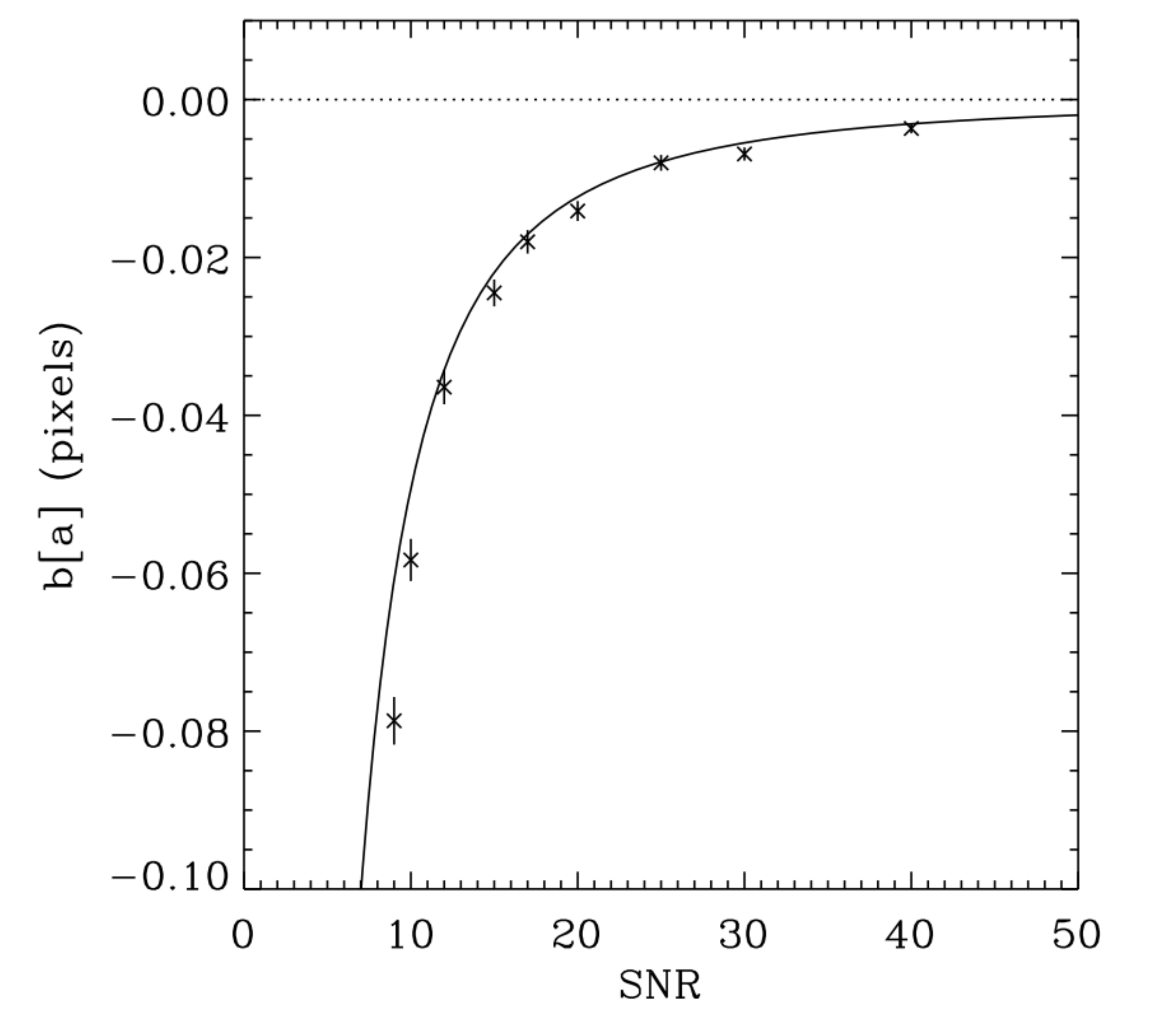}
\includegraphics[width=70mm]{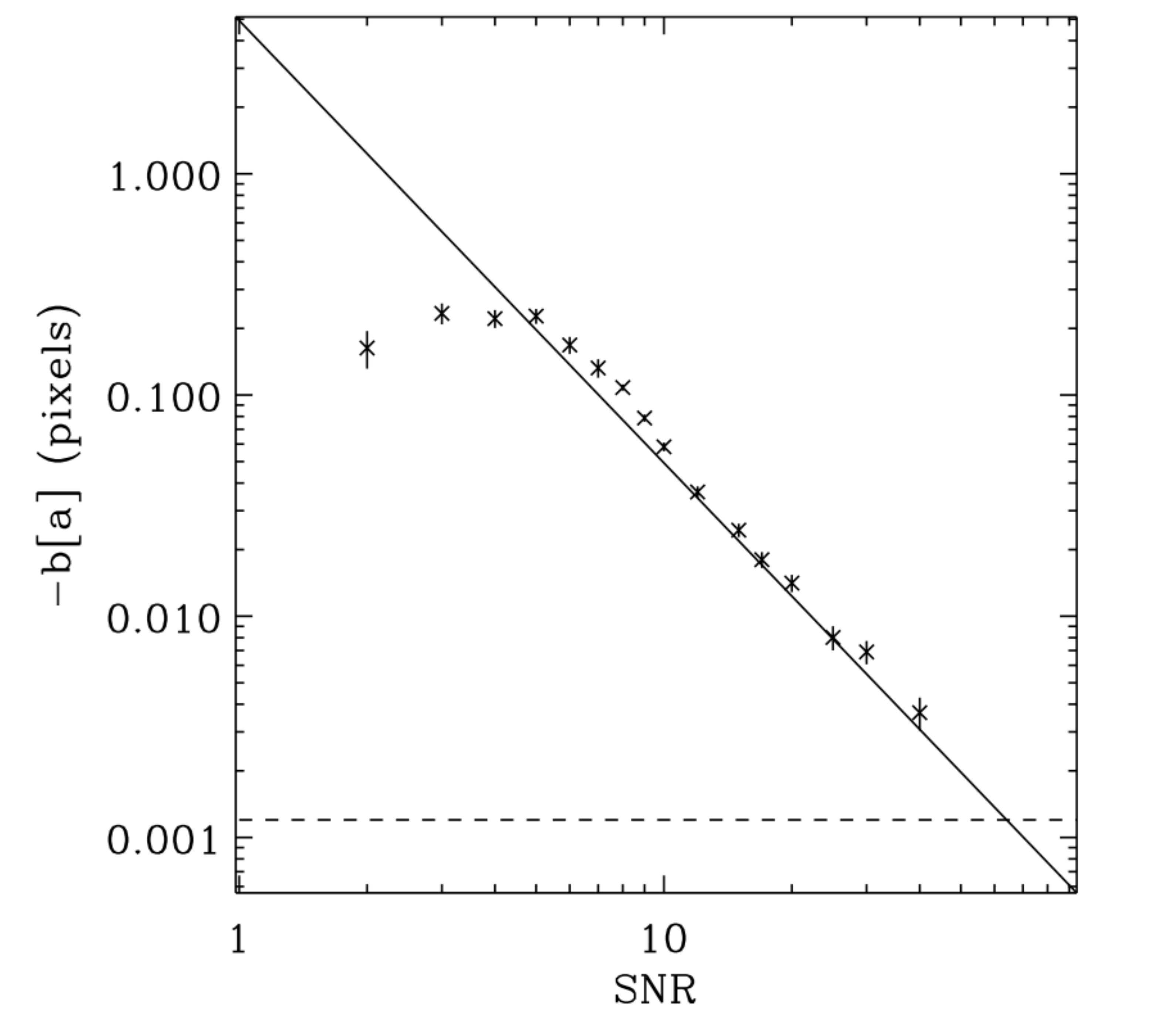}
\caption{Bias $b[a]$ of the size estimator $a$ as a function of ${\rm
    SNR}=\rho[f_0]$ in linear {(upper panel)} and log {(lower panel)}
  axes. In both panels, the solid line correspond to the analytical
  model while the squares were derived from simulations of repeated
  experiments. The horizontal dashed line in the bottom panel
  corresponds to the requirement for future all sky surveys (see
  text).}
\label{fig:bias_a}
\end{center}
\end{figure}

Figure~\ref{fig:bias_a} shows the dependence of the bias $b[a]$ on the
SNR $\rho[f_0]$. Again, our expression in Equation~\ref{eq:bias} is a
good approximation for SNR $\ga 10$, with deviations below this value
{likely} due to higher order terms. The horizontal, dashed line in the
{lower} panel corresponds to the requirement for future all sky
surveys $b[a]/a \simeq \delta \gamma \simeq 3\times 10^{-4}$ as
discussed in \S\ref{properties} and in \citet{Amara:2007as}. The bias
for galaxies with SNR $\sim 10$ in this toy model is $b[a]/a \simeq
0.015$, which is nearly two order of magnitude {greater} than this
requirement and comparable to the expected weak lensing signal $\delta
\gamma \simeq 0.02$.

\section{Elliptical Gaussian model}
\label{elliptical}
We now consider a slightly more realistic model of a galaxy consisting
of a 2D elliptical Gaussian galaxy with 6 parameters. This is the
smallest number of parameters needed to measure galaxy shapes for weak
lensing in practice as they correspond to the 2 centroid coordinates,
flux, major and minor axes and position angle of the galaxy. In
practice the models for galaxies are typically non-Gaussian and more
complicated in order to {better} describe realistic galaxies. The
6-parameter Gaussian model is nevertheless useful to study the
behaviour of the bias in the multi-parameter case.

Let us thus consider a galaxy surface brightness given by a 2D
elliptical Gaussian (without PSF convolution) that we parametrise as
(see also \citealp{paulinetal08} for a slightly different
parametrisation):
\begin{equation}
  f({\mathbf x};{\mathbf a})= \frac{A}{2\pi \sqrt{a_1 a_2}} \exp \left[ -\frac{1}{2} ({\mathbf x}-{\mathbf x}^a)^{T} {\mathbf A}^{-1} ({\mathbf x}-{\mathbf x}^a) \right],
\label{eq:elliptical_gaussian}
\end{equation}
where $a_1$ and $a_2$ are the (rms) major and minor axes of the
Gaussian respectively, $A$ is a parameter which {determines} the
amplitude, ${\mathbf x}^a$ is the centroid, and $^T$ {denotes} the
transpose operator.  The quadrupole moment matrix ${\mathbf A}$ is a
$2\times2$ symmetric matrix which can be written as
\begin{equation}
{\mathbf A}={\mathbf R}(\alpha)^{T} 
\left( \begin{array}{cc}  a_1^2 & 0 \\ 0 & a_2^2 \end{array}  \right)
{\mathbf R}(\alpha),
\label{eq:a_matrix}
\end{equation}
where $\alpha$ is the position angle of the major axis
counter-clockwise from the $x$-axis and 
\begin{equation}
{\mathbf R}(\alpha)= \left( 
\begin{array}{cc} \cos \alpha & \sin \alpha \\ -\sin \alpha & \cos \alpha \end{array}
\right)
\label{eq:rot}
\end{equation}
is the rotation matrix which aligns the coordinate system with the
major axis. In Appendix~\ref{app:elliptical}, we show that the
amplitude, centroid, and quadrupole moment matrix are simply related
to the multipole {moments} of galaxy surface brightness.

Let us consider the MLEs for this 2-dimensional Gaussian with the
following 6 free parameters
\begin{equation}
{\mathbf a}=\left (x_1^a,x_2^a,A,a_1,a_2,\alpha \right).
\label{eq:6_params}
\end{equation}
In the continuum limit (Eq.~\ref{eq:continuous}), the Fisher matrix
equations~(\ref{eq:fisher}) can be computed analytically. This is
facilitated by rotating into the coordinate system aligned with the
major and minor axes of the galaxy, before performing the integral of
the surface of the galaxy. The resulting Fisher matrix $F_{ij}$ and
covariance matrix ${\rm cov}[a_i,a_i]$ of the parameters are given in
Equation~\ref{eq:cov_elliptical} in Appendix~\ref{app:elliptical}.
{We note} that, with this parametrisation, the Fisher matrix is
conveniently nearly diagonal\footnote{Note that in the parametrisation
  of \citet{paulinetal08} the Fisher matrix is also not quite
  diagonal, due to covariance terms between their rotated centroid and
  the position angle neglected in this paper.}.  The corresponding rms
statistical errors $\sigma[a_i]={\rm cov}[a_i,a_i]^{1/2}$ are listed
in the third column in the top part of
Table~\ref{tab:elliptical_gaussian}, to leading order in the {SNR} of
the amplitude $A$, {defined as}
\begin{equation}
\rho[A]\equiv \frac{A}{\sigma [A]}= \frac{A}{\sqrt{4\pi} h \sigma_n}.
\label{eq:snr_a}
\end{equation}

Using {Equation}~(\ref{eq:bias}) after the same change of coordinates
and some cumbersome algebra, we can also derive the bias in the model
parameters which are given in Equation~\ref{eq:bias_elliptical} and
listed in the last column of Table~\ref{tab:elliptical_gaussian}. Note
that several of the parameters have a singularity as the galaxy
becomes circular, i.e. when $\epsilon = 0$, which occurs at
$a_1=a_2$. This follows from the fact that, in this limit, the
position angle $\alpha$ becomes degenerate.

From these expressions, and using
{Equations}~(\ref{eq:g_cov}-\ref{eq:g_bias}), we can derive the error
and bias for derived quantities. The {lower} part of
Table~\ref{tab:elliptical_gaussian} provides the definition,
statistical errors and biases for several commonly used parameters
such as the flux $F^{(0)}$ (see also Equation~\ref{eq:flux}), the
average radius $a=\sqrt{a_1^2+a_2^2}$ and two definitions of the
ellipticities $\epsilon$ and $e$. All these parameters are biased to
second order in $\rho[A]$. Interestingly, the average radius $a$ is
biased in the elliptical case even in the absence of a PSF
convolution, as a result of covariances with other parameters. Also,
it is interesting to note that the rms error and bias of $a_1$ and
$a_2$ have a singularity in the circular case, but that the
singularity cancels when they are combined to form the mean radius
$a$.  We can also verify that the scaling of
Equation~\ref{eq:scalings} for the variance and bias of the parameters
{holds} up to multiplicative factors of order unity.

\begin{table*}
\centering
\begin{minipage}{120mm}
  \caption{Statistical errors and biases of parameters for 6-parameter elliptical Gaussian model. Errors and biases are shown to leading order in $\rho$
    \label{tab:elliptical_gaussian}}
  \begin{tabular}{@{}llll@{}}
  \hline \hline
   Parameter & Symbol & Stat. Error & Bias \\  
   Name  & $a$   &     $(\sigma[a]/a)\rho[A]$      & $(b[a]/a) \rho[A]^{2}$ \\
  \hline \hline
\multicolumn{4}{c}{Model Parameters} \\
\hline
Centroid  & $x^{a}_{1}$ &  $\sqrt{2(a_1^2 \cos^2 \alpha + a_2^2 \sin^2 \alpha)}/x_1^a$ & 0  \\
Centroid  & $x^{a}_{2}$ &  $\sqrt{2(a_2^2 \cos^2 \alpha + a_1^2 \sin^2 \alpha)}/x_2^a$ & 0  \\
Amplitude &A & 1 & $5/2$\\
Major axis &$a_1$ & $\sqrt{2}$ & $\epsilon^{-1}$ \\
Minor axis &$a_2$ & $\sqrt{2}$ & $-\epsilon^{-1}$ \\
Position angle &$\alpha$ & $2 \sqrt{\epsilon^{-2}-1} / \alpha$ & 0\\
\hline
\multicolumn{4}{c}{Derived Parameters} \\
\hline
Flux & $F^{0}=A\sqrt{a_1 a_2}$ & $\sqrt{2}$ & 5/2 \\
Radius mean  & $a=\sqrt{a_1^2+a_2^2}$ & $\sqrt{1+\epsilon^2}$ &  1 \\
Ellipticity (quadratic) & $\epsilon=\frac{a_1^2-a_2^2}{a_1^2+a_2^2}$ & $2 (1-\epsilon^2)/\epsilon$ & 
$2 (1-\epsilon^2)/\epsilon^2$ \\
Ellipticity (linear) & $e=\frac{a_1-a_2}{a_1+a_2}$ & $(1-e^2)/e$ & $(1-e^4)/(2e^2)$ \\
\hline
\hline
\end{tabular}
\end{minipage}
\end{table*}

\section{Conclusions}
\label{conclusions}

In this paper we have studied the effect of noise bias on {MLEs} for
weak lensing shape parameters in the presence of additive Gaussian
noise.  We have derived general expressions for the covariance and
bias for {ML-estimated} parameters of 2D galaxy images, which are
given by Equations~\ref{eq:covariance} and \ref{eq:bias}. The bias
vanishes for linear models, but is generally non-zero for models which
are non-linear in the parameters and depend on the model
parametrisation.  To illustrate the effect of the noise bias we have
{calculated} analytical expressions for the variance and bias for a
toy model consisting of a 2D circular Gaussian galaxy, convolved with
a circular Gaussian PSF, with the galaxy size as a single free
parameter.  We have compared these predictions with {careful}
numerical simulations and found them to be in good agreement.  We also
{provide} analytical results for a 2D elliptical Gaussian with
6-parameters.

We find that the variance and bias of the parameters are generically
of order $\rho^{-2}$, where $\rho$ is the SNR of the galaxies.  For
galaxies with $\rho \sim 10$, which is typical of weak lensing
surveys, this implies a bias in the parameters of order $\rho^{-2}
\sim 0.01$. {This is} comparable to the weak lensing shear signal
$\gamma \sim 0.02$ and nearly two orders of magnitude {greater} than
the {systematic} shear {tolerance} required for future all sky
surveys, $\delta \gamma \sim 3\times 10^{-4}$.  {Although} derived
using the {specific case of the MLE} in the presence of Gaussian
noise, our results are likely to be generic {across a number of
  measurement techniques}.  {This may} contribute towards explaining
why current weak lensing surveys are limited by systematics, and {why
  finding} sufficiently accurate methods has been difficult.

To solve this problem, the following ways forward are possible
\begin{itemize}
\item Use higher SNR galaxies. This is an obvious solution but it is
  costly in practice as it leads to a sharp drop in the useful surface
  density of galaxies, or {requires} longer exposure times and/or
  larger telescopes.
\item Avoid non-linearities, either by choosing linear models or using
  moment-based methods. This was the idea behind shapelets methods,
  but {we} note that in all cases some level of non-{linearity} is
  unavoidable as the centroid and size (of the basis functions or
  weight function) are intrinsically non-linear.
\item Go beyond ML estimation by using, e.g., Bayesian methods or
  other averaging techniques. For instance, the introduction of
  stacking methods was a breakthrough in the GREAT08 challenge
  \citep{GREAT08mnras}.
\item Calibrate the bias. While order-by-order correction methods
  exist for the MLEs, the models used in practice to model galaxies
  (e.g., exponential, de Vaucouleur, Sersic, or a multi-component
  combination) convolved with observed PSFs will be complex and thus
  not offer analytical expressions for the bias. Instead, numerical
  simulations will be needed for the bias calibration. This is the
  approach described in the accompanying paper by Kacprzak et
  al. (2012).
\end{itemize}

\bigskip We thank Gary Bernstein, Julien Carron, Jo\"{e}l Berg\'{e},
St\'{e}phane Paulin-Henriksson and Lisa Voigt for helpful
discussions. Sarah Bridle acknowledges support from the Royal Society
in the form of a University Research Fellowship, and both Sarah Bridle
and Barnaby Rowe acknowledge support from the European Research
Council in the form of a Starting Grant with number 240672.  Part of
Barnaby Rowe's work was done at the Jet Propulsion Laboratory,
California Institute of Technoloy, under contract with NASA.

\bibliographystyle{mn2e}
\bibliography{mybib}{}

\appendix

\section{Bias for general MLE}
\label{app:general}
In this Appendix, we provide the derivation of the main results for
the variance and bias of a general MLE of parameters given in
\S\ref{general} for additive, uncorrelated Gaussian noise. For the
model describe in \S\ref{general} , the likelihood is
\begin{equation}
L \propto e^{-\chi^2/2}
\end{equation}
where
\begin{equation}
\chi^2({\mathbf a})= \sum_{p} \sigma_{n}^{-2} \left[ f({\mathbf x}_p;{\mathbf a}^{t}) 
+ n({\mathbf x}_p) - f({\mathbf x}_p;{\mathbf a}) \right]^{2}.
\label{eq:chi2_ext}
\end{equation}
The MLE $\hat{\mathbf a}$ is then defined as the value of the
parameters ${\mathbf a}$ which maximises the likelihood $L$ or,
equivalently, which minimises $\chi^2$, i.e.\ for which
\begin{equation}
\left. \frac{\partial \chi^2}{\partial {\mathbf a}} \right|_{\hat{\mathbf a}} =0.
\label{eq:dchi2da}
\end{equation}

To proceed, we expand this expression in terms of the inverse of the
{SNR}, $\rho \sim f/n$, of the object. This can be conveniently done
by rewritting $n({\mathbf x}_p) \rightarrow \alpha n({\mathbf x}_p)$
and $\hat{\mathbf a} = {\mathbf a}^{t} + \alpha \delta {\mathbf
  a}^{(1)}+\alpha^2 \delta {\mathbf a}^{(2)} +\cdot \cdot \cdot $,
where $\alpha$ is a dimensionless order parameter which scales as
$\alpha \sim \rho^{-1}$.  We then Taylor expand $f(x,a)$ about
$a^{t}$, collect like powers of $\alpha$ in Equation~\ref{eq:dchi2da},
and set $\alpha=1$. The terms of order $\alpha$ yield
\begin{equation}
  \delta a_{i}^{(1)} = (F^{-1})_{ij} \sum_{p} \sigma_n^{-2} n({\mathbf x}_p) \frac{\partial f({\mathbf x}_p;{\mathbf a})}{\partial a_j}.
\end{equation}
where the fisher matrix $F_{ij}$ was defined in
Equation~\ref{eq:fisher}. {Taking} the average of this expression and
using the fact that the noise is unbiased, i.e.\ $\langle n({\mathbf
  x}_p) \rangle=0$, we see that the estimator is unbiased to this
order. Taking the average of the product $\langle \delta a_{i}^{(1)}
\delta a_{j}^{(1)} \rangle$ and using the fact that the noise is
uncorrelated, i.e.\ $\langle n({\mathbf x}_{p}) n({\mathbf x}_{p'})
\rangle= \delta_{p,p'} \sigma^2_n$, we obtain the expression for the
covariance of the parameters to leading order given in
Equation~\ref{eq:covariance}.

The terms of order $\alpha^2$ yield
\begin{equation}
\delta a_{i}^{(2)} = -\frac{1}{2}(F^{-1})_{ij} (F^{-1})_{kl} B_{jkl}, 
\end{equation}
and thus {gives} Equation~\ref{eq:bias} for the bias to {leading}
order, where the bias tensor was defined in
Equation~\ref{eq:bias_tensor}.

These results can also be derived from the general expressions of the
Fisher Matrix
\begin{equation}
F_{ij}=\left \langle - \frac{\partial^2 \ln L}{\partial a_i \partial a_j} \right \rangle
\end{equation}
and for the bias tensor for the MLE 
\begin{equation}
B_{ijk}=\left \langle -  \frac{1}{2}\frac{\partial^3 \ln L}{\partial a_i \partial a_j \partial a_k} 
+ \frac{\partial \ln L}{\partial a_j } \frac{\partial^2 \ln
  L}{\partial a_i \partial a_k} \right \rangle {.}
\end{equation}

In the limit of small pixels, the sum over pixels in the expressions
above can be approximated by a continuous integral
\begin{equation}
\label{eq:continuous}
\sum_p \simeq \int \frac{d^2 x}{h^2},~~~ {\rm as}~~ h \rightarrow 0,
\end{equation}
where $h$ is the pixel size.

\section{Results for the elliptical Gaussian model}
\label{app:elliptical}
In this {Appendix}, we provide results for the elliptical Gaussian
model defined in Equation~\ref{eq:elliptical_gaussian} as a function
of the 6 parameters listed in Equation~\ref{eq:6_params} in the
presence of additive, {uncorrelated} Gaussian noise as defined in
Equation~\ref{eq:f_obs}.

With this parametrisation, the flux, or {zeroth} order moment of the
Gaussian is given by
\begin{equation}
\label{eq:flux}
F^{(0)}=\int d^2x~f({\mathbf x};{\mathbf a}) = A \sqrt{a_1 a_2}.
\end{equation}
The centroid, or first order moments are given by
\begin{equation}
\frac{F_i^{(1)}}{F^{(0)}}=\frac{1}{F^{(0)}} \int d^2x~x_i f({\mathbf x};{\mathbf a}) = x_i^a,
\end{equation}
and the quadrupole moment matrix, containing the second order
moments, is given by
\begin{equation}
  \frac{F_{ij}^{(2)}}{F^{(0)}}=\frac{1}{F^{(0)}} \int d^2x~x_i x_j f({\mathbf x};{\mathbf a}) = A_{ij}.
\end{equation}

For this model, in the continuum limit (Eq.~\ref{eq:continuous}), the
Fisher Matrix $F_{ij}$ (Eq.~\ref{eq:fisher}) can be derived
analytically by rotating into a coordinate system {with axes} parallel
to the major and minor axes of the galaxy (using the rotation matrix
in Eq.~\ref{eq:rot}) before performing the integrals over the surface
of the galaxy.  The covariance matrix of the parameters is then
obtained by inverting the Fisher matrix (Eq.~\ref{eq:covariance})
which yields
\begin{eqnarray}
\label{eq:cov_elliptical}
{\rm cov}[\hat{a}_i,\hat{a}_j] & = & \left(
\begin{array}{cccccc}
2A_{11} & 2A_{12} & 0 & 0 & 0 & 0 \\
2A_{12} & 2A_{22} & 0 & 0 & 0 & 0 \\
0 & 0 & A^{2} & 0 & 0 & 0 \\
0 & 0 & 0 & 2a_1^2 & 0 & 0 \\
0 & 0 & 0 & 0 & 2a_2^2 & 0 \\
0 & 0 & 0 & 0 & 0 & \frac{4 a_1^2 a_2^2}{(a_1^2-a_2^2)^2}
\end{array}
\right) \nonumber \\
&  & \times \rho[A]^{-2} + O(\rho^{-3}),
\end{eqnarray}
where $A_{ij}$ are the components of the quadrupole moment matrix
${\mathbf A}$ defined in Equation~\ref{eq:a_matrix}, and $\rho[A]$ is
the SNR of the amplitude $A$ given in Equation~\ref{eq:snr_a}.

The bias in the parameters can be derived using Equation~\ref{eq:bias}
which yields
\begin{equation}
\frac{b[a_i]}{a_i}=  \left(0,0,\frac{5}{2},\epsilon^{-1},-\epsilon^{-1},0 \right) \rho[A]^{-2} + O(\rho^{-3}),
\label{eq:bias_elliptical}
\end{equation}
where the quadratic ellipticity $\epsilon$ is defined in Table~\ref{tab:elliptical_gaussian} .

The resulting rms errors and biases for the model parameters are
summarised in the {upper} part of Table~\ref{tab:elliptical_gaussian}.

\end{document}